\documentclass[superscriptaddress,twocolumn,prd,preprintnumbers,amsmath,amssymb,nofootinbib]{revtex4}

\usepackage{graphicx}
\usepackage{dcolumn}
\usepackage{bm}
\usepackage{epsfig}
\usepackage{amssymb}
\usepackage{amsmath}

\newcommand{\lsim}{\mathrel{\hbox{\rlap{\lower.55ex\hbox{$\sim$}} \kern-.3em \raise.4ex \hbox{$<$}}}}
\newcommand{\gsim}{\mathrel{\hbox{\rlap{\lower.55ex\hbox{$\sim$}} \kern-.3em \raise.4ex \hbox{$>$}}}}
\newcommand{\beq}{\begin{equation}}
\newcommand{\eeq}{\end{equation}}
\newcommand{\beqa}{\begin{eqnarray}}
\newcommand{\eeqa}{\end{eqnarray}}
\newcommand{\drm}{\mathrm{d}}

\newcommand{\apjs}{ApJ Supp.}
\newcommand{\jcap}{JCAP}
\newcommand{\aj}{Astronom. J}
\newcommand{\apjl}{ApJL}
\newcommand{\mnras}{MNRAS}
\newcommand{\aap}{A\&A}
\newcommand{\physrep}{Phys. Rep.}
\newcommand{\plb}{Phys. Lett. B}

\newcommand{\sigv}{\langle \sigma v \rangle}

\newcommand{\Om}{\Omega_\mathrm{M}}

\newcommand{\msun}{M_\odot}

\newcommand{\deltamin}{\delta_\mathrm{min}}
\newcommand{\sigmahor}{\sigma_{\mathrm{hor}}}
\newcommand{\zeq}{z_\mathrm{eq}}
\newcommand{\feq}{f_\mathrm{eq}}
\newcommand{\deltam}{M_i}
\newcommand{\Smin}{S_\mathrm{min}}
\newcommand{\uas}{\,\mu\mathrm{as}}
\newcommand{\mdm}{m_\chi}

\begin{document}
\title{A new probe of the small-scale primordial power spectrum: astrometric microlensing by ultracompact minihalos}
\author{Fangda Li}\email{fangda.li@utoronto.ca}
\affiliation{Department of Physics, University of Toronto, 60 St.~George Street, Toronto, Ontario M5S 3H8, Canada}
\author{Adrienne L. Erickcek}\email{erickcek@cita.utoronto.ca}
\affiliation{Canadian Institute for Theoretical Astrophysics, University of Toronto, 60 St.~George Street, Toronto, Ontario M5S 3H8, Canada}
\affiliation{Perimeter Institute for Theoretical Physics, 31 Caroline St. N, Waterloo, Ontario N2L 2Y5, Canada}
\author{Nicholas M. Law}\email{law@di.utoronto.ca}
\affiliation{Dunlap Institute for Astronomy and Astrophysics, University of Toronto, 50 St.~George Street, Toronto, Ontario M5S 3H8, Canada}
\date{\today}

\begin{abstract}
The dark matter enclosed in a density perturbation with a large initial amplitude ($\delta \rho/\rho \gtrsim 10^{-3}$) collapses shortly after recombination and forms an ultracompact minihalo (UCMH).  Their high central densities make UCMHs especially suitable for detection via astrometric microlensing: as the UCMH moves, it changes the apparent position of background stars.  A UCMH with a mass larger than a few solar masses can produce a distinctive astrometric microlensing signal that is detectable by the space astrometry mission Gaia.  If Gaia does not detect gravitational lensing by any UCMHs, then it establishes an upper limit on their abundance and constrains the amplitude of the primordial power spectrum for $k\!\sim\!2700$ Mpc$^{-1}$.   
These constraints complement the upper bound on the amplitude of the primordial power spectrum derived from limits on gamma-ray emission from UCMHs because the astrometric microlensing signal produced by an UCMH is maximized if the dark-matter annihilation rate is too low to affect the UCMH's density profile.  
If dark matter annihilation within UCMHs is not detectable, a search for UCMHs by Gaia could constrain the amplitude of the primordial power spectrum to be less than 10$^{-5}$; this bound is three orders of magnitude stronger than the bound derived from the absence of primordial black holes.  
\end{abstract}

\maketitle

\section{Introduction}

Structure formation is hierarchical: small-scale density perturbations form the first dark matter minihalos, and these minihalos are later absorbed into larger dark matter structures.   If these small-scale density perturbations have the same average initial amplitude $(\delta \rho/\rho \simeq 10^{-5})$ as the large-scale density perturbations probed by the cosmic microwave background (CMB) and observations of large-scale structure, then the first dark matter minihalos form long after the Universe became matter-dominated, at redshifts $z\lsim60$ \cite{DMS05, IME10}.  Density perturbations with larger initial amplitudes can form dark matter halos much earlier; if $(\delta \rho/\rho \simeq 10^{-3})$ when the perturbation enters the Hubble horizon, then the dark matter within the overdense region will collapse to form a minihalo shortly after recombination ($z\simeq1000$) \cite{RG09}, and larger fluctuations can form minihalos while the Universe is still radiation dominated \cite{BDEK10}.  Since these minihalos form in a denser environment than later-forming minihalos, they have high central densities, and they have been dubbed ultracompact minihalos (UCMHs).  

The abundance of UCMHs measures the primordial power spectrum of density fluctuations because UCMHs form from density perturbations with large initial amplitudes.  The upper limit on the amplitude of the primordial power spectrum derived from the absence of UCMHs is extremely valuable because UCMHs probe the primordial power spectrum on scales far smaller than those accessible via observations of the CMB and large-scale structure.  As detailed below, UCMHs offer substantial improvements over other bounds on the amplitude of the small-scale power spectrum, but all existing limits on the abundance of UCMHs are derived from their potential emission of dark-matter annihilation products \cite{SS09, JG10, Zhang11, BSA11}.  In this paper we present a new method of searching for UCMHs that can provide strong constraints on the amplitude of the primordial power spectrum even if dark matter does not self-annihilate. 

On cosmological scales, the CMB and large-scale structure provide a direct probe of the primordial power spectrum.  The amplitude of temperature fluctuations in the CMB is proportional to the amplitude of the primordial curvature power spectrum on wavelengths of \mbox{$\sim\!\!10^4$ Mpc} to \mbox{$\sim\!20$ Mpc} \cite{CBI03, ACBAR09, Quad09, WMAP7yrPower, ACT11}.  On these scales, the CMB indicates that the primordial power spectrum of curvature perturbations is nearly scale-invariant with an amplitude of $\sim\!\!\!2\times 10^{-9}$ \cite{NC09, BPVV11}.  Measurements of the matter power spectrum inferred from observations of large-scale structure \cite{SDSSDR7, VKB09, TSW11, STA11} and weak gravitational lensing \cite{HYG02, vWMH05, HMvW06, HSS07, BHS07} probe similar scales as the CMB, while observations of the Lyman-$\alpha$ forest reach slightly smaller wavelengths ($\sim\!\!2$ Mpc) \cite{MSB06}.  All these observations are also consistent with a nearly scale-invariant spectrum of primordial fluctuations with an amplitude of $\sim\!2\times10^{-9}$ \cite{TZ02, ACT11, BPVV11}.  

These measurements of the primordial power spectrum are often cited as evidence for inflation \cite{Guth80, AS82, Linde82} because many inflationary scenarios predict that the primordial power spectrum should be nearly scale-invariant over a wide range of scales (see Ref. \cite{LLK97} for a review).  However, the power spectrum deviates from scale-invariance in several inflationary models.  Features in the inflationary potential, such as a steps, kinks, or bumps, enhance the power spectrum at specific scales \cite{1989PhRvD..40.1753S, 1992JETPL..55..489S, 1994PhRvD..50.7173I, 1998GrCo....4S..88S, 2008PhRvD..77b3514J}.  Multifield inflationary models can also produce steps and other features in the power spectrum \cite{1987PhRvD..35..419S, Polarski:1992dq, 1997NuPhB.503..405A,  2011JCAP...01..030A, 2012JCAP...05..008C}.  Interactions between the inflaton and other fields generate deviations from scale-invariance \cite{1988PhLB..214..508K}:  particle production during inflation produces a bump in the power spectrum \cite{2000PhRvD..62d3508C, 2009PhRvD..80l6018B, 2010PhRvD..82j6009B};  the power spectrum oscillates if the inflaton loses energy during inflation \cite{2009JCAP...02..014A}; and excess power on small scales is generated if the inflaton has an axial coupling to a gauge field \cite{2011PhRvL.106r1301B, 2011arXiv1110.3327B}.  Several other inflationary models \cite{1996NuPhB.472..377R, 1998PhRvD..58f3508C, 2000PhRvD..61h3518M, 2001PhRvD..63l3501M, 2010JCAP...09..007B}, including running-mass inflation \cite{1997PhRvD..56.2019S, 1999PhRvD..59f3515C, 1999PhRvD..60b3509C} and hybrid inflation \cite{2011JCAP...03..028G, 2011JCAP...07..035L,2011JCAP...11..028B}, also predict enhanced perturbations on small scales.  Small-scale perturbations can also be significantly amplified after inflation by the QCD phase transition \cite{1999PhRvD..59d3517S} or a nonstandard thermal history \cite{ES11}.

Clearly, it is essential that we measure the amplitude of the power spectrum on small scales.  Other than UCMHs, the only probes of density perturbations with wavelengths smaller than $\sim$1 Mpc are spectral distortions in the CMB and primordial black holes (PBHs).   Upper bounds on the deviation of the CMB spectrum from that of a perfect black-body constrain the integrated amplitude of the power spectrum for wavelengths between 0.6 kpc and 1 Mpc because the energy released by the dissipation of these perturbations does not completely thermalize \cite{1994ApJ...430L...5H,2012A&A...540A.124K, 2012arXiv1202.0057C,JEB12}.  However, there are several mechanisms capable of generating CMB spectral distortions \cite{Chluba2011therm}, so a future detection would not necessarily provide information about the primordial power spectrum.  

Like UCMHs, PBHs directly probe the amplitude of the primordial power spectrum; PBHs form when density fluctuations with an initial amplitude of $\delta \rho/\rho \gsim 0.3$ enter the horizon \cite{Carr75, NJ99}.  The abundance of PBHs is tightly constrained over a wide range of PBH masses; these constraints imply an upper bound on the amplitude of the primordial curvature power spectrum of 0.01-0.06 for wavelengths between 600 Mpc and $10^{-16}$ pc \cite{JGM09}.   Limits on the abundance of UCMHs can provide far more powerful constraints on the amplitude of the primordial power spectrum because less extreme density perturbations are required to form UCMHs than are necessary to form PBHs.

If the dark matter is thermally produced in the early universe, then it self-annihilates \cite{SWO88, GG91, JKG96}, and the high density of dark matter within UCMHs enhances the annihilation rate.  It was quickly recognized that the Large Area Telescope on the Fermi Gamma-Ray Space Telescope (Fermi-LAT) \cite{Atwood2009} could detect UCMHs as gamma-ray sources \cite{SS09} and that measurements of the diffuse gamma-ray background constrain the abundance of UCMHs \cite{BDEK10, LB10, YFH11}.  Since UCMHs form shortly after recombination, gamma-ray emission from UCMHs can also have a profound effect on the ionization history of the Universe \cite{Zhang11, YHCZ11, YCL11}.  Consequently, measurements of the optical depth to the surface of last scattering also limit the fraction of dark matter that may be contained in UCMHs.  Finally, the fact that Fermi-LAT has not yet detected gamma-rays from dark matter annihilation within an UCMH puts a strong upper bound on their number density within the Milky Way, which implies an upper bound on the amplitude of the primordial power spectrum; this analysis was initially proposed by Ref. \cite{JG10} and was recently refined and extended by Ref. \cite{BSA11}.   If dark matter self-annihilates, then these limits lower the upper bound on the primordial power spectrum by several orders of magnitude; if the mass of the dark matter particle is less than 1 TeV and it is a standard thermal relic, then UCMH abundance constraints imply that the amplitude of the primordial curvature power spectrum does not exceed $2\times10^{-6}$ - $1.5\times10^{-7}$ on wavelengths between 1 Mpc and 0.3 pc \cite{BSA11}.

Unfortunately, the numerous searches for dark matter annihilation products have not yet detected any evidence that dark matter self-annihilates \cite{2010JCAP...01..031S, 2010JCAP...05..025A, 2010JCAP...11..041A, Abdo:2010dk, 2010PhRvD..82l3503E, Ackermann:2011wa, Acciari:2010pja, Ripken:2010ja, Aleksic:2011jx}.
There may be an asymmetry between dark matter particles and antiparticles (see e.g.  \cite{Dodelson:1991iv, KLZ09, DMS10, BR11}),
or the cross section for dark matter annihilations may be much lower than the value expected for a thermal relic in the standard scenario \cite{GKR01, GG06}.
The only guaranteed signatures of UCMHs are gravitational; to obtain model-independent constraints on the amplitude of primordial power spectrum, we must search for UCMHs via their gravitational effects.  

UCMHs can be detected through photometric microlensing \cite{RG09}.  When an UCMH passes in front of a star, it produces a light curve that is similar to the light curve produced by a compact object (point lens) with a mass equal to the mass enclosed in the UCMH's Einstein radius (assuming that the core radius of the UCMH is much smaller than its Einstein radius).  Currently, photometric microlensing searches tell us that point-like objects with masses between $10^{-6} \,\msun$ and $10 \,\msun$ contain less than $3\%$ - $10\%$ of the dark matter within the Milky Way halo \cite{Tisserand:2006zx, OGLE11a, OGLE11b}.  However, despite its compactness, only a small fraction of a UCMH's mass is enclosed within its Einstein radius.  If the lensed source is located in the Large Magellanic Cloud, a UCMH with less than $10 \,\msun$ within its Einstein radius has more than 98\% of its mass \emph{outside} this radius at its formation, and this percentage increases as the UCMH accretes more matter.  Therefore, photometric microlensing searches do not limit the abundance of UCMHs because these constraints are satisfied even if all dark matter immediately collapses into UCMHs at the time of matter-radiation equality.

In this paper, we consider astrometric microlensing by UCMHs, and we evaluate the constraints on the primordial power spectrum that could be obtained from a search for UCMHs using high-precision astrometry.   When an UCMH passes in front of a star, the location of that star's image will move as the angular separation between it and the UCMH changes.   The image trajectory produced by a diffuse lens is easily distinguished from the astrometric microlensing signature of a point mass, offering a distinctive way to detect dark matter substructures.  In Ref.~\cite{EL11}, Erickcek and Law explored possibility of detecting astrometric microlensing by conventional dark matter subhalos; unfortunately, the number density of dark matter subhalos predicted by numerical simulations of galaxy-sized dark matter halos is too small for a blind search for astrometric microlensing events by subhalos to be successful.  However, Ref. \cite{EL11} noted that the probability of detecting astrometric microlensing by subhalos is greatly enhanced if the subhalos are more compact and more numerous than predicted by numerical simulations.  Since UCMHs may be abundant in our galaxy and have higher central densities and steeper density profiles than typical, later-forming minihalos \cite{RG09}, they are optimal targets for detection by astrometric microlensing.  Nevertheless, the image deflections produced by UCMHs are measured in microarcseconds, and high-precision astrometry is required to detect their astrometric microlensing signatures.

Recent years have seen great growth in high-precision astrometric measurement capabilities using both ground and space-based instruments. Current all-sky astrometric catalogs derived from ground-based observations achieve tens-of-milliacrsecond precisions (e.g. \cite{Ivezic2008, Zacharias2010}), while from space the Hipparcos satellite \cite{Perryman1997} provided milliarcsecond-precision astrometry for stars brighter than 9th magnitude across the entire sky.\footnote{JMAPS \cite{Hennessy2010} will provide similar performance.} For small fields, ground-based and space-based long-term astrometric monitoring campaigns can achieve milliarcsecond-precision astrometry for much fainter objects (e.g. \cite{Miniak2009, Pravdo2006, Henry2009,Subasavage2009,Monet2010,Anglada2011}). The development of adaptive optics (AO) systems has boosted the optical ground-based astrometric precision still further for some targets; large telescopes can achieve 100-microarcsecond precision on single targets \cite{Cameron2009, Lu2009} while new laser-guide-star systems designed for smaller telescopes (in particular, the Robo-AO system \cite{Law2009, Baranec2009, Law2012}) offer the opportunity to cover large numbers of targets with similar precisions.  Multi-conjugate AO systems for large telescopes offer the possibility of this performance over much larger fields (e.g. \cite{Meyer2011, Rochau2011, Hart2010, Neichel2010}). Optical and radio interferometry have demonstrated still higher astrometric precisions (e.g. \cite{vanBelle2008, Muterspaugh2010, Vincent2011}) but these techniques generally require optically or radio-bright sources.

Although these capabilities would be very useful for follow-up, a sensitive search for UCMHs requires all-sky coverage at the tens-of-microarcsecond level. Fortunately, the Gaia mission, planned for launch in 2013, offers these capabilities \cite{Lindegren2011}.  The satellite contains two 1.45m$\times$0.5m mirrors imaged onto a common focal plane containing a 1-gigapixel astrometric camera, precision spectrophotometers and a radial velocity spectrograph. The instrument is designed to achieve end-of-mission astrometric accuracies of 5-14 microarcseconds for stars brighter than $12^{\mathrm{th}}$ magnitude ($\sim\!\!25$ microarcseconds for $15^{\mathrm{th}}$ magnitude stars). The important metric for astrometric transient searches is the single-epoch measurement precision; for Gaia, this corresponds to a sky-averaged measurement precision of 23-microarcseconds for the $\sim\!\!5\times10^6$ stars \cite{2005ESASP.576..163D} brighter than $12^{\mathrm{th}}$ magnitude in the Gaia passband.\footnote{at the time of the Gaia Critical Design Review (performed in April 2011) -- www.rssd.esa.int/index.php?project
=Gaia\&page=Science\_Performance}

Inspired by the Gaia mission, we evaluate the constraints on the UCMH abundance that could be obtained from monthly observations of 5 million stars over six years, and we use these constraints to forecast bounds on the amplitude of the primordial power spectrum.  We begin in Section \ref{sec:UCMHprop} by reviewing the properties of UCMHs with particular attention to their density profiles.  In Section \ref{sec:AMD}, we present the image trajectories produced by astrometric microlensing by UCMHs, and we show how these image trajectories depend on the UCMH density profile.  We then describe our detection strategy and define lensing cross sections: the area of the sky surrounding an UCMH in which a star would be detectably lensed.  We use these lensing cross sections to calculate the probability that surveys with varying levels of astrometric precision would detect astrometric microlensing by UCMHs.  In Section \ref{sec:powerspectrum}, we calculate the constraints on the UCMH abundance that could be obtained by high-precision astrometric surveys, and we translate these constraints to upper bounds on the amplitude of the primordial power spectrum on small scales.  These bounds complement the bounds obtained from gamma-ray searches for UCMHs, and this connection is explored in Section \ref{sec:gammarays}.  Finally, we summarize and discuss our results in Section \ref{sec:summary}.

\section{Properties of UCMHs}
\label{sec:UCMHprop}
UCMHs form when over-dense regions with \mbox{$\delta \rho/\rho_0 \gsim 10^{-3}$} enter the Hubble horizon.  
The initial mass of a UCMH, $\deltam$, equals the entire dark matter content of this over-dense region at horizon entry; if $R$ is the comoving radius of the overdensity and $a$ is the scale factor, then the over-density enters the horizon when $aR=H^{-1}$, and 
\beqa
\deltam&=& \frac{4\pi}{3} \rho_{\chi,\mathrm{hor}} (a_\mathrm{hor}R)^3 = \frac{H_0^2}{2G}\Omega_\chi R^3,\\
&=& 130\, \msun \left(\frac{\Omega_\chi h^2}{0.112}\right) \left(\frac{R}{\mathrm{kpc}}\right)^3
\label{delm}
\eeqa
where $H_0= 100h$ km s$^{-1}$ Mpc$^{-1}$ is the Hubble constant, and $\Omega_\chi$ the present-day ratio of the dark matter density to the critical density.  Throughout this paper, the subscript ``hor" indicates that the quantity is to be evaluated at horizon entry.

UCMHs do not grow significantly prior to the redshift of matter-radiation equality ($z_\mathrm{eq} \simeq 3250$) \cite{MOR07}.  After matter-radiation equality, an isolated UCMH grows linearly with the scale factor: 
\begin{equation}
M_h(z) = \deltam\left(\frac{1+\zeq}{1+z} \right).
\end{equation}
Previous work \cite{SS09, JG10} has adopted $z = 10$ as the redshift after which hierarchical structure formation prevents further accretion, leading to a UCMH mass growth factor
\begin{equation}
g \equiv \frac{M_h(z = 0)}{M_h(z_{\mathrm{eq}})} \simeq 300.
\end{equation}  
However, this much accretion is only possible if there is enough free dark matter to be accreted for each minihalo to grow by $g$.  If a fraction $f_\mathrm{eq}$ of the dark matter is contained in UCMHs at matter-radiation equality, then $\feq\lsim 1/300$ is required for $g=300$.  
For larger values of $f_\mathrm{eq}$, $g$ is limited by the availability of free dark matter; in general, $g = \mathrm{min} \lbrace {f_\mathrm{eq}}^{-1}, (1+\zeq)/{11} \simeq 300 \rbrace$. 
We note that, if UCMHs are sufficiently numerous to accrete all the dark matter at high redshift, it is likely that UCMHs will interact with other UCMHs before they are absorbed in larger dark matter halos.  We assume that these interactions do not affect the central region of the UCMHs' density profiles; we revisit this assumption in Section \ref{sec:powerspectrum}.

The theory of self-similar secondary infall \cite{Bert85, FG84, RG09} models the radial accretion of collisionless particles from a uniform background onto a point mass and predicts a power-law UCMH density profile $\rho (r) \propto r^{-9/4}$.  However, the angular momentum of infalling dark matter particles in any realistic cosmological context will be non-negligible \cite{ROM08} and will lead to a significantly shallower profile in the central regions of these minihalos.  Dark matter annihilation can also limit the maximum central density of UCMHs.  Since the dynamic astrometric microlensing signal is strongly dependent on the radial index of the density profile \cite{EL11}, we conservatively assume that both of these mechanisms lead to a constant density core with radius $r_c$.

We first consider the case of a core resulting from non-radial infall.  Assuming that $f_\mathrm{eq} < 1/300$, the truncation radius of a UCMH is given by \cite{Ric07, MOR07}
\begin{equation}
r_t(z) = 0.019 \left( \frac{1000}{1+z} \right) \left[ \frac{M_h (z)}{\msun} \right]^{1/3} \mathrm{pc}.
\label{truncationradius}
\end{equation}
In the absence of a core, the density profile of a UCMH with initial mass $\deltam$ is
\begin{equation}
\begin{split}
\rho (r) &= \frac {3M_h(z)}{16 \pi \, r_t(z)^{3/4} \, r^{9/4}} \\
&= 1.6 \times 10^7 \left(\frac{\deltam}{\msun}\right)^{0.75}\left(\frac{r}{\mathrm{10^{-3} \, pc}} \right)^{-2.25} \msun \, \mathrm{pc^{-3}}.
\end{split}
\label{uncoredprofile}
\end{equation}
The density profile is time-independent because the material accreted after the UCMH forms has a high radial velocity when it reaches the inner regions of the halo and does not contribute significantly to the time-averaged density in those regions \cite{Bert85}.  

When the UCMH forms, the dark matter has a small velocity dispersion; thus the particles do not fall into the UCMH on perfectly radial orbits.  If we assume that the dark matter velocity dispersion at UCMH formation is negligibly affected by the presence of the UCMHs themselves, then an infalling dark matter particle's initial tangential velocity at a radius $r_t$ from the center of the UCMH is  \cite{ROM08}
\begin{equation}
\sigma_{\bot} \simeq \frac{2}{\pi} \, 1.9 \times 10^{-4} \, \left(\frac{1+z_\sigma}{1000} \right)^{-0.78} \left(\frac{\deltam}{\msun} \right)^{0.28} \mathrm{km} \, \mathrm{s^{-1}}.
\end{equation}
The radial infall approximation breaks down when the tangential velocity of a particle falling from $r_t$ exceeds the local Keplerian orbital velocity.  Using the density profile in Eqn.~(\ref{uncoredprofile}) and conservation of angular momentum, this occurs at a radius
\begin{equation}
r_{\mathrm{c, nr}} \simeq \left[ \frac{\sigma_{\bot}^8 r_t^{11}}{G^4 M_h^4} \right]^{1/7}.
\end{equation}
We take $r_{\mathrm{c, nr}}$ to be the radius of the constant density core caused by non-radial infall.  The radius of the non-radial infall core can be expressed solely in terms of the redshift of evaluation $z_\sigma$:
\begin{equation}
r_{\mathrm{c, nr}} = 1.5 \times 10^{-6} \,\left(\frac{1+z_\sigma}{1000}\right)^{-2.41} \left(\frac{\deltam}{\msun} \right)^{0.272} \mathrm{pc}.
\end{equation}
This radius is usually evaluated at $z_\sigma$ of 1000, which corresponds to the redshift at which perturbations of amplitude $10^{-3}$ have grown to reach the critical overdensity for collapse and the UCMH forms \cite{RG09}.

We now consider cores produced by the annihilation of dark matter.  The maximum central density of an ultracompact minihalo of age t is
\begin{equation}
\begin{split}
\rho_{\mathrm{max}} & = \frac{m_\chi}{\langle \sigma v \rangle \, t} \\
&= 2.0 \times 10^{8} \left(\frac{m_{\chi}}{100 \, \mathrm{GeV}} \right) \times \\
& \left( \frac{\langle \sigma v \rangle}{3 \times 10^{-26} \, \mathrm{cm^3/s}}\right)^{-1}\left( \frac{t}{13.7 \, \mathrm{Gyr}}\right)^{-1} \msun \, \mathrm{pc^{-3}}
\end{split}
\end{equation}
where $m_{\chi}$ is the mass of the dark matter particle and $\langle\sigma v\rangle$ is the product of the annihilation cross section and relative DM particle velocity \cite{SS09}.  The radius of the annihilation core is equal to the radius at which the UCMH density reaches $\rho_{\mathrm{max}}$; from Eq.~(\ref{uncoredprofile}) we see that 
\begin{equation}
\begin{split}
r_{\mathrm{c, ann}} &= 3.3 \times 10^{-4} \left(\frac{\deltam}{\msun} \right)^{1/3} \left(\frac{m_{\chi}}{100 \, \mathrm{GeV}} \right)^{-4/9} \times \\
& \left( \frac{\langle \sigma v \rangle}{3 \times 10^{-26} \, \mathrm{cm^3/s}}\right)^{4/9}\left( \frac{t}{13.7 \, \mathrm{Gyr}}\right)^{4/9} \mathrm{pc}.
\end{split}
\end{equation}

The prevailing core radius in the UCMH density profile is simply the greater of $r_{\mathrm{c, nr}}$ and $r_{\mathrm{c, ann}}$.  The final cored UCMH density profile we consider for microlensing is
\begin{equation}
\rho(r) = \rho_0\left( 1+ \frac{r}{r_c}\right)^{-9/4}.
\label{coredprofile}
\end{equation}
We set the density to zero for $r > r_t$.  If dark matter annihilation is the prevailing factor in determining the core radius ($r_{\mathrm{c, ann}} > r_{\mathrm{c, nr}}$) then $\rho_0$ is simply the maximum central density of the UCMH $\rho_\mathrm{max}$.  If the core due to non-radial infall prevails, then we first determine the overall normalization factor $\rho_{0, \mathrm{nr}}$ such that the the mass of the halo at matter-radiation equality $M_i$ is equal to the volume integral of this density profile within the truncation radius at $z_\mathrm{eq}$, \emph{assuming that there is no annihilation}.  In some cases, $\rho_{0, \mathrm{nr}}$ may be greater than the maximum density allowed by annihilation.  So in the case where $r_{\mathrm{c, nr}} > r_{\mathrm{c, ann}}$, we take $\rho_0$ to be the lesser of $\rho_{0, \mathrm{nr}}$ and $\rho_\mathrm{max}$.

\section{Astrometric microlensing by UCMHs}
\label{sec:AMD}

\subsection{Image Trajectories}
\label{sec:AMD1}

In gravitational microlensing by a spherically symmetric thin lens, the apparent image of a background star is deflected from its true position by an angle
\begin{equation}
\vec{\alpha} = \frac{d_{\mathrm{ls}}}{d_{{s}}}\left[\frac{4GM_{\mathrm{2D}}(\xi)}{c^2\xi} \right] \hat{\xi},
\label{deflectionangle}
\end{equation}
where $d_{\mathrm{ls}}$ is the distance between the lens and source planes, $d_{{s}}$ is the distance between the observer and source planes, $\vec{\xi}$ is the impact parameter of the light ray in the lens plane with respect to the center of the lens, and $\hat{\xi} \equiv \vec{\xi}/\xi$ points from the lens center to the star.  $M_{\mathrm{2D}}(\xi)$ is the mass enclosed by a cylinder of radius $\xi$ collinear to the optical axis.  Using Eq.~(\ref{coredprofile}), it is given by
\begin{equation}
M_{\mathrm{2D}}(\xi) = 4 \pi \int_{0}^{\xi} \int_{0}^{\sqrt{r_t^2-(\xi^\prime)^2}}{\!\!\!\! \frac{\rho_0 \, \xi^\prime}{ \left[1+\frac{\sqrt{(\xi^\prime)^2+z^2}}{r_c} \right]^{9/4}} \, dz \, d\xi^\prime}.
\end{equation}
Since we are only interested in lensing events in the Galactic halo, we can take $d_\mathrm{ls}/d_s = 1-(d_l/d_s)$, where $d_l$ is the distance from the observer to the lens plane.  In the weak microlensing regime, the angular separation between the true image of the star and the center of the lens $\beta$ is much greater than the deflection angle $\alpha$.  Since these two angles are always collinear, we can approximate $\vec{\xi} = d_l(\vec{\alpha} + \vec{\beta})\simeq d_l \vec{\beta}$.  In this case, Eq.~(\ref{deflectionangle}) becomes a simple function for $\vec{\alpha}(\vec{\beta})$.

\begin{figure}
\resizebox{\columnwidth}!{
\includegraphics{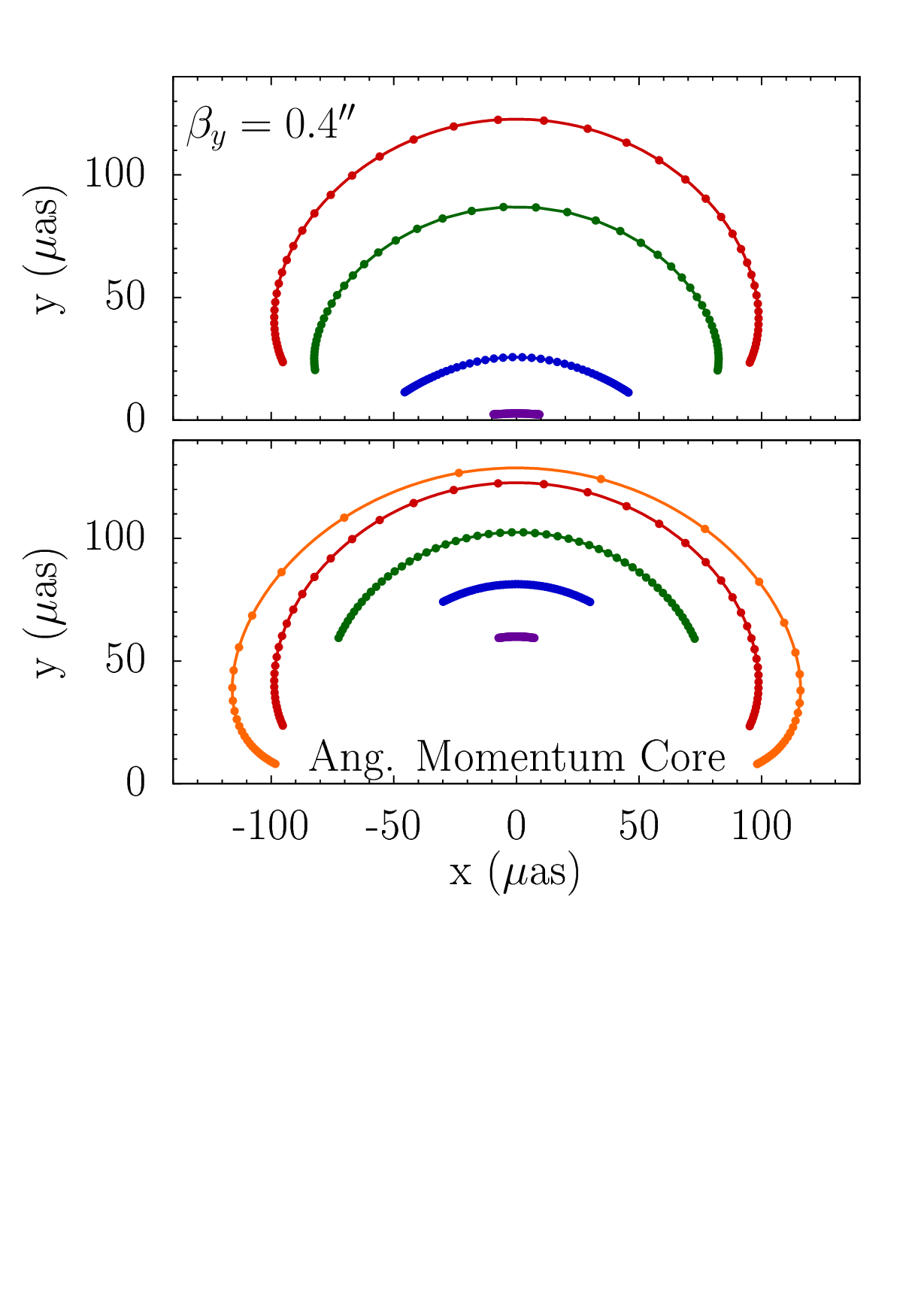}}
\caption{Astrometric microlensing trajectories for a star at $5 \, \mathrm{kpc}$ and a UCMH at $50 \, \mathrm{pc}$ with a tangential velocity of $200 \, \mathrm{km \, s^{-1}}$.  The initial mass of the UCMH is $\deltam = 11 \msun$.  The true position of the star is at the origin and the UCMH is moving horizontally below the star; as the UCMH moves from left to right, the apparent image of the star moves from right to left.   The four 
years surrounding the moment of closest approach are shown, with points plotted every 27 days.  The upper plot shows the effect of varying core size while fixing the impact parameter at $0.4$ arcseconds.  The highest trajectory corresponds to a core caused by the angular momentum of infalling material only, while the lower three correspond to annihilation cores with $m_\chi = 100 \, \mathrm{GeV}$ and $\langle \sigma_v \rangle = \lbrace 3 \times 10^{-30}, 3 \times 10^{-28}, 3 \times 10^{-26} \rbrace \, \mathrm{cm^3 \, s^{-1}}$.  These correspond to core sizes $\lbrace2.8 \times 10^{-6},1.2 \times 10^{-5}, 9.3 \times 10^{-5}, 7.2 \times 10^{-4}\rbrace$ parsecs.  The lower plot shows the effect of varying impact parameter with a fixed angular momentum core for the same UCMH configuration.  In descending order, the trajectories correspond to impact parameters of $\beta_y = \lbrace 0.1, 0.4, 1, 4, 10\rbrace$ arcseconds.}
\label{trajectorydiffcores}
\end{figure}

Figure \ref{trajectorydiffcores} shows the trajectories of the apparent images of a star at $5 \, \mathrm{kpc}$ for varying core sizes and impact parameters.  We choose the co-ordinate system such that the x-axis is parallel to the proper motion of the UCMH and the true position of the star is at the origin.  Here, the separation angle $\beta$ can be decomposed into an impact parameter $\beta_y$ that is constant throughout the lensing event, and a perpendicular component $\beta_x$ along which all the lens motion during the event occurs.  Four years of the lensing event are shown in Fig.~\ref{trajectorydiffcores}, with points are plotted every 27 days.  In the orientation shown, the UCMH starts at the left and moves right, while the image of the star starts right of its true position and eventually moves left.  When $\beta$ is still large, the image motion of the star is slow; the image accelerates until it reaches its maximum velocity when the UCMH is directly below the star, precisely when when the magnitude $\beta$ is minimized with $\beta_x = 0$.

It is easily seen from Eq.~(\ref{deflectionangle}) that more massive haloes generate a larger overall lensing trajectory.  However, the presence of a larger constant density core in such haloes also changes the trajectories' shapes compared to those of their smaller, cuspier counterparts.  As shown in the upper plot of Fig.\ref{trajectorydiffcores}, the magnitude of the astrometric deflection $\alpha$, its velocity, and its acceleration are all diminished as $r_c$ increases if the halo mass is fixed.  The presence of a core dramatically reduces the lensing mass $M_\mathrm{2D}$ when $\xi \lsim r_c$, but the relative effect of the core on $M_\mathrm{2D}$ is smaller for larger $\xi$.  Thus the vertical extent of the trajectory, which is determined when $\xi$ is small, is more highly suppressed than its horizontal extent, which is determined when $\xi$ is large.  This leads to a flattened lensing trajectory compared to that of a smaller core.

Given a particular UCMH, large impact parameters $\beta_y$ pose two challenges for astrometric microlensing detection.  First, increasing the impact parameter beyond $r_c$ entails an overall reduction in the amplitude of the image's vertical deflection.  Second, since we do not know the true position of the star, any detection scheme is highly dependent on the velocity of the apparent image, which is also reduced as $\beta_y$ increases.  This is shown in the lower plot of Fig.\ref{trajectorydiffcores}, where the size of the core is fixed but the impact parameter varies.  As $\beta_y$ increases, the deflection angle is reduced (for $\beta_y d_l \gtrsim r_c$) and image motion slows dramatically - similar to the effect of increasing the core size.  Consequently, any detection scheme for lensing events is strongly biased to those with small impact parameters.

\subsection{Detection Strategy and Lensing Cross Section}

\begin{figure*} 
\resizebox{\textwidth}!{
\includegraphics{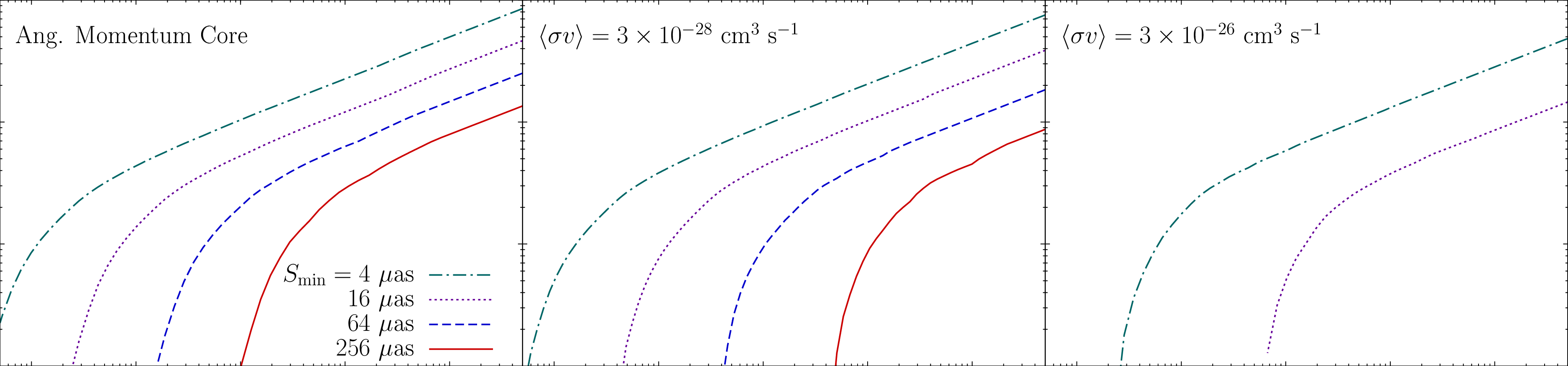}}
\caption{Lensing cross sections in square arcseconds as a function of initial UCMH mass $M_i$ with $d_l = 50\, \mathrm{pc}$ and $d_s = 2\,\mathrm{kpc}$.  The four curves correspond to different values of $S_\mathrm{min}$.  The three panels show the constraints for three different UCMH density profiles.  In the left panel, the core radius of the UCMH is given by the breakdown of radial infall, while in the right two panels, the core radius is determined by dark matter annihilation.  The dark matter particle mass is assumed to be 100 GeV, and the core radius is determined by the ratio $\langle \sigma v\rangle/m_\chi$.}
\label{crosssections}
\end{figure*}

We use the same detection technique as in Ref.~\cite{EL11}.  We assume a 6 year mission lifetime with 82 epochs per target: a typical Gaia observing scenario.  As the true position of the lensed star is unknown, we must search for anomalies in its proper motion.  The first 2 years are used as a calibration period, after which we fit the observed star positions for the effects of parallax and proper motion.  The lensing signal $S$ is then simply the difference between the star's extrapolated position from the fit and the lensed position:

\begin{equation}
S = \sqrt{\sum_{i = 1}^{N_\mathrm{epochs}} \left[ (X_{l, i}-X_{p, i})^2+(Y_{l, i}- Y_{p, i})^2 \right]}
\end{equation}
where $N_\mathrm{epochs}$ is the number of epochs and $\left( X_{l, i}, \, Y_{l, i} \right)$ and $\left( X_{p, i}, \, Y_{p, i} \right)$ are the lensed position of the star and the predicted position of the star at each epoch, respectively.  

To calculate the signal, we establish a Cartesian co-ordinate system on the sky with normalized impact parameter $\tilde{\beta} \equiv \beta_y d_l / (v_\mathrm{t} \, t_\mathrm{obs})$ on one axis and phase, defined $\varphi \equiv \beta_\mathrm{x, 0} d_l/(v_\mathrm{t} \, t_\mathrm{obs})$, on the other.  Here, $\beta_\mathrm{x, 0}$ is the $\beta_x$ co-ordinate of the UCMH at the start of the observational period, $v_t$ is the transverse halo velocity, and $t_\mathrm{obs}$ is the length of the observational period.  Throughout our analysis, we have $t_\mathrm{obs} = 4$ years, the difference between the total mission length and the calibration period.  

The significance of the phase $\varphi$ is that it denotes the proportion of the observational period that passes \emph{before} the angular separation between the star and lens reaches its minimum ($\beta_x = 0$).  Each point on the $\tilde{\beta} - \varphi$ plane corresponds to a particular initial star-lens geometric configuration for which we calculate the lensing signal $S$ as defined above.

A comprehensive analysis would incorporate a probability distribution for $v_t$, taking into consideration the velocity distribution of UCMHs.  However, as a first approximation, we note that UCMHs are expected to have random motions with respect to the rest frame of the Galactic halo.  Thus we expect our relative velocity with respect to typical UCMH to be approximately equal to the velocity of the Sun with respect to the dark matter halo \mbox{$v_t \simeq v_\odot = 200 \, \mathrm{km\, s^{-1}}$}.  The remainder of our results assume this value for $v_t$.  The effects of different transverse velocities on the lensing signal will be discussed at the end of this section.

The lensing cross section for a single UCMH $A_{\mathrm{lensed}}$ is the area on the sky surrounding the halo within which a lensed star's signal would exceed the statistically significant minimum for detection $S_{\mathrm{min}}$, determined by the astrometric instrument employed.  From numerical simulations of the detection technique \cite{EL11}, $S_\mathrm{min} = 1.47 \, \mathrm{SNR} \, \sigma_{\mathrm{inst}}$, where $\mathrm{SNR}$ is the desired signal-to-noise ratio and $\sigma_{\mathrm{inst}}$ is the single-epoch instrumental astrometric uncertainty.  Given a particular UCMH and minimum signal $S_\mathrm{min}$, it is a simple matter to determine the area on the $\tilde{\beta} - \varphi$ plane that generates $S > S_\mathrm{min}$ and convert this to a solid angle on the sky $A_\mathrm{lensed}$.  

We also consider a reverse calibration method that could be implemented given a complete 6-year dataset from an astrometric survey.  The time-reversed motion of the stars during last two years are used to calibrate a reverse proper motion fit, which is then used to detect lensing events during the first four years.  This process allows us to detect lensing events whose signals are largely generated during initial calibration period.  In practice, the contribution to the lensing cross section from lensing events with phases $0.5 \leq \varphi \leq 1$ is effectively doubled, neatly accounting for those lensing events for which the lens passes the star during the first two years of observation.

We restrict our lensing cross section to only include stars that reach their point of closest approach to the UCMH's center during the 4 year observational period ($0 \leq \varphi \leq 1$), allowing us to distinguish UCMH microlensing events from sources on binary orbits by the distinctive acceleration patterns in their astrometric trajectories.  We note that a more sophisticated detection technique could potentially be subject to a weaker restriction.  Furthermore, if $\beta_y \gsim r_t$ then $M_{2D}(\xi)$ is constant, making the lensing event undistinguishable from that induced by a point mass.  Therefore we also restrict our UCMH lensing cross section within a circle of radius $r_t$ in the lens plane.  In practice, however, the cross-sections for the values of $\Smin$ and $M_i$ considered almost always fall well inside of $r_t$. 

Individual UCMH lensing cross-sections are shown in Fig.~\ref{crosssections}.  Given $\langle \sigma v \rangle/m_\chi$ and $S_\mathrm{min}$, there is a minimum UCMH mass that generates the required signal to be detected.  The area increases rapidly as $M_i$ increases beyond this minimum before settling into a weaker power-law dependence.  Fixing $S_\mathrm{min}$, an increase in the core size leads to a greater required halo mass for detection.  Finally, a reduction in $\Smin$ increases the lensed area of any given halo.

We now return to the issue of the UCMH velocity.  In Ref.~ \cite{EL11}, it was found that for subhalos with a coreless, untruncated power-law density profiles, the $v_t$ dependence could be factored out of the expression for the lensing cross section using the dimensionless parameters $\varphi$ and $\tilde{\beta}$, leading to a simple scaling of $A_\mathrm{lensed} \propto v_t^2$ since $v_t t_\mathrm{obs}$ was the only length scale in the system.  We find that the same scaling with $v_t$ also applies for UCMH lensing over a wide range of $M_i$ despite the introduction of additional length scales $r_c$ and $r_t$.  However, for extremely small UCMHs ($M_i \lsim 10^{-2} \msun$), large transverse velocities can cause the total distance traversed by the halo during the observational period $v_t t_{\mathrm{obs}}$ to exceed $r_t$.  In this case, for some duration of the observational period, the light ray's impact parameter $\xi$ in the lens plane is greater than $r_t$ and the enclosed mass $M_{2D}(\xi)$ remains constant.  Similar to the case when the impact parameter is greater than the truncation radius $\beta_y d_l \gsim r_t$, the lensing trajectory will then tend to close in a fashion identical to that of a point lens, making the UCMH lensing event difficult to distinguish.  Our rejection of all lensing scenarios with $\xi > r_t$ at any point in the observation or calibration period leads to a deviation from the $A_\mathrm{lensed} \propto v_t^2$ scaling for small UCMHs; if $M_i \lsim 10^{-2} \msun$, increasing $v_t$ decreases the lensing cross-section. 

\subsection{False positives} 

The lensed stars' trajectories are near-ellipsoidal on very long
timescales, with the stars returning to their original positions
(within $\sim\!\! 10 \uas$) after a few hundred years (for $\deltam =5\,\msun$) to tens of thousands of years (for $\deltam =1000\,\msun$).  
In our relatively short six-year observation window we can only see a
segment of the total trajectory.  The fastest-moving part of image's trajectory
necessarily occurs during the observation period; away from that time the motion is very small (see
Fig.~\ref{trajectorydiffcores}).  Consequently, the trajectories induced by UCMHs are very different from the
near-circular astrometric trajectories induced by point-source
microlensing; these trajectories return to the star's true position much more quickly. (We refer readers to Ref.~\cite{EL11} for a detailed
discussion of this and other possible false positives.)  Because of its
range of possible trajectories, orbital motion is the most important
false positive for astrometric motion produced by UCMHs.  However,
most orbital motion can be immediately distinguished by either its
repetition (for the shorter periods), or by the production of an
anomalous acceleration during the calibration period.

We conducted extensive Monte-Carlo simulations to explore the types of
orbits that could mimic the lensing trajectories. If the detection
strategy rejects sources because of an anomalous acceleration during
the calibration period, we found that no Keplerian orbit was capable
of producing a false positive signal (for eccentricities $<$0.99; see
Ref.~\cite{EL11} for details). Longer-period highly eccentric orbits
were rejected because of a detectable signal during the calibration
period. Shorter-period highly eccentric orbits could produce a small
enough acceleration to escape notice during calibration, but had
orbital periods that would lead to a repeated motion within the Gaia
mission.

All microlensing-produced trajectories also produce some motion during
the calibration period, but only the events in which the star and the UCMH reach their closest point early in the observational period (i.e. events with small phases) produce calibration-period accelerations that are detectable by Gaia. The threshold for
calibration-period rejection of a target should be set after an optimization of the number of rejected false-positives compared to
ignored microlensing events. As part of an improved detection strategy, which we leave for future work, this optimization could include further
methods of false-positive rejection such as radial velocities, probabilistic assessments based on the proper motion and parallax of
the target, and further ground-based astrometric monitoring of the most promising cases.

Keplerian orbit false-positives could also be distinguished by Gaia
directly, without the astrometric calibration period. Keplerian
orbital motion, likely produced by a central, bright star being moved
by a distant fainter companion, has attendant radial velocity
variations. Using the distance to the target star,  a typical value
for its mass, the size of its apparent astrometric motion, and the
apparent period of the motion for a typical UCMH-like signal, we can
estimate the size and orbital radius of a companion star that is
necessary to produce the astrometric motion. In almost all cases we
find that the companion would have to be a low-mass brown dwarf. Using
the estimated masses and orbital periods, we find that all astrometric
orbits that could mimic a lensing trajectory produce radial
velocities changes in excess of 0.75 km s$^{-1}$ on Gaia mission lifetime
timescales. Even without an astrometric calibration period, many of
possible false-positive orbits can thus be directly rejected by Gaia's
Radial Velocity Spectrometer, which will achieve km/s precision on
these bright targets. 

Finally, a detected UCMH lensing event would provide a measurement of the direction and speed of the UCMH's motion on the sky. As discussed in Ref.~\cite{EL11}, this information could be used to predict the lensing of stars along the UCMH's projected trajectory.  Subsequent observations of these stars using ground-based telescopes could then provide confirmation that an UCMH was responsible for the detected event.

\subsection{Lensing Probabilities}

\begin{figure*}
\resizebox{\textwidth}!{
\includegraphics{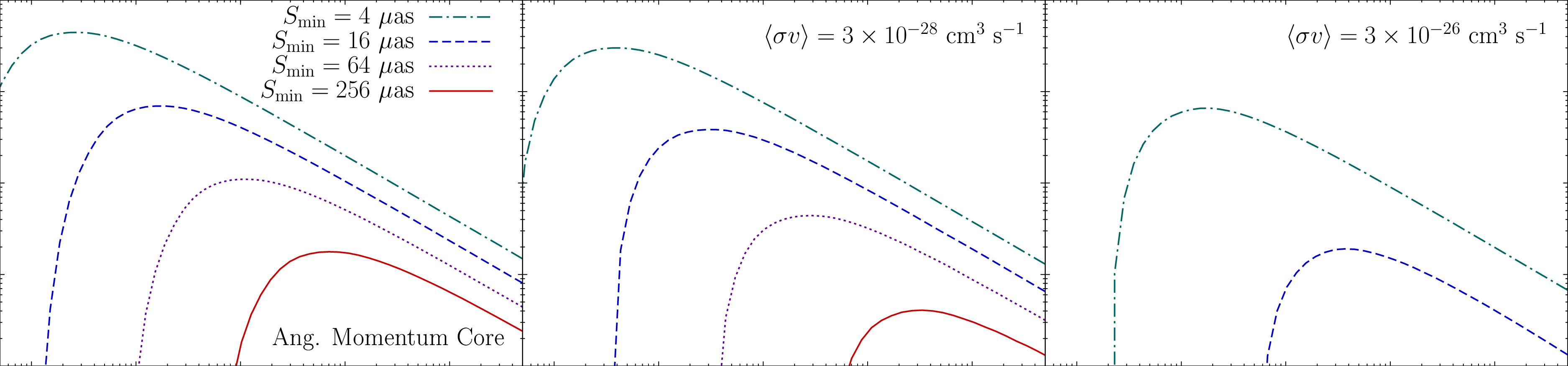}}
\caption{Lensing probabilities as a function of initial UCMH mass $M_i$, assuming that $\feq = 1$ and the observed stars have an average distance $d_s = 2 \, \mathrm{kpc}$.  As in Fig.\ref{crosssections}, the four lines correspond to different values of $S_\mathrm{min}$, and the three panels show the constraints for three different UCMH density profiles.}
\label{lens_prob}
\end{figure*}

To determine the likelihood of observing a lensing event, one must first calculate the total lensed solid angle in the sky $A_\mathrm{total}$.  This is equal to the sum of the individual $A_\mathrm{lensed}$ over all UCMHs, if they are rare and small enough that one can assume that they do not overlap.  If UCMHs originate from a strongly localized enhancement in the primordial power spectrum, they have the same initial mass $M_i$.  Taking their growth factors $g$ to be equal, we can extend this assumption to the current mass $M_h$.  

We assume that UCMHs are homogeneously distributed with respect to the dark matter density $\rho_\mathrm{dm}$.  If the fraction of dark matter in UCMHs today ($f_0$) is 1 (occurring for $f_\mathrm{eq} \gsim 1/300$), then $n_\mathrm{ucmh} = f_\mathrm{eq} \rho_\mathrm{dm}/M_i$, where $\rho_\mathrm{dm}$ is the dark matter density.  Otherwise, $n_\mathrm{ucmh} = f_0 \rho_\mathrm{dm}/M_h \simeq f_0 \rho_\mathrm{dm}/(300 M_i)$.  We then have 
\begin{equation}
A_{\mathrm{total}} = 4 \pi \int_{0}^{d_s} \, n_{\mathrm{ucmh}}(d_l) A_{\mathrm{lensed}} (d_l) \, d_l^2 \, d(d_l).
\label{totalarea}
\end{equation}

The relation between $A_\mathrm{lensed}$ and $d_l$ is complicated since the lens distance $d_l$ enters the microlensing equation (\ref{deflectionangle}) in $d_\mathrm{ls}$ and $\vec{\xi}$.  When all other lensing parameters $M_i, \tilde{\beta}, \varphi, v_t, t_\mathrm{obs}$ are held fixed, the resulting lensing signal is a complex function of $d_l$.  Thus given some $S_\mathrm{min}$, $A_\mathrm{lensed}(d_l) \, d_l^2$ is also a complex function of $d_l$ (however, it is nearly proportional to $1-d_l/d_s$ for most interesting values of $M_i$ and $S_\mathrm{min}$).  When evaluating Eq.~(\ref{totalarea}), we compute $A_\mathrm{lensed}(d_l) \, d_l^2$ at discrete values of $d_l$ and then integrate over an interpolation based on these points.

We take $d_s = 2 \, \mathrm{kpc}$ as an estimate of the average source distance in an astrometric survey like Gaia, which conservatively represents its full discovery space including the large number of distant giant stars at the brightness levels we target.  The relation $A_\mathrm{lensed}(d_l) d_l^2\propto 1-d_l/d_s$ implies that $A_\mathrm{total}$ is approximately linear in $d_s$.  Thus our estimate effectively averages over a homogeneous distribution of lensing sources centered around 2 kpc. 

To simplify the calculation of Eq.~(\ref{totalarea}), we note that, assuming a Navarro-Frenk-White (NFW) profile for the dark matter halo of the Milky Way, the solar orbital radius $\simeq 8 \,  \mathrm{kpc}$ lies well within the scale radius $r_s \sim 20 \, \mathrm{kpc}$ \cite{XRZ08, BHM05, Klypin02}, which implies that the dark matter density $\rho_\mathrm{dm}$ scales locally as $1/r$.  Taking $\rho_\mathrm{dm, 0}$ to be the local dark matter density, $\rho_\mathrm{dm}$ varies over a sphere of radius $d_s = 2 \, \mathrm{kpc}$ centered on the Sun from $(4/3) \rho_\mathrm{dm, 0}$ in the region closest to the Galactic center to $(4/5) \rho_\mathrm{dm, 0}$ in the outermost region.  The relation $A_\mathrm{lensed}(d_l) d_l^2\propto 1-d_l/d_s$ further suppresses the impact of deviations from $\rho_\mathrm{dm, 0}$ because the cross sections of nearby UCMHs dominate $A_\mathrm{tot}$.  Assuming that $\rho_\mathrm{dm} \propto 1/r$, and $A_\mathrm{lensed}(d_l) d_l^2\propto 1-d_l/d_s$, the contribution to $A_\mathrm{tot}$ varies by only $\pm 10\%$ when comparing opposing lines of sight directly towards and away from the Galactic center, with all other lines of sight falling between these two extremes.  Consequently, we can approximate $\rho_\mathrm{dm} = \rho_\mathrm{dm, 0}$ within the entire sphere so that $n_{\mathrm{ucmh}}(d_l)$ can be factored out of the integral in Eq.~(\ref{totalarea}).

However, the local dark matter density is uncertain \cite{CU10, GRL11, WB10}, with estimates ranging from 0.1 to $1.3 \, \mathrm{GeV \, cm^{-3}}$ and uncertainties from $7 \%$ to a factor of 3 \cite{SNG10}.  In our analysis, we take the conventional value $\rho_\mathrm{dm} = 0.4$ GeV cm$^{-3}$.  The total lensing cross-section is linear in $n_\mathrm{ucmh}$, and it will be shown when we consider the UCMH survival rate in Sec. IV that changing $A_\mathrm{total}$ by a factor of order unity does not greatly affect our final bounds on the primordial curvature perturbation.

The probability that any given star in the sky at $d_s$ is being lensed by a UCMH is $P_{\mathrm{lensed}} = A_{\mathrm{total}}/A_{\mathrm{sky}}$, where $A_{\mathrm{sky}} \simeq 5 \times 10^{11} \, \mathrm{as^2}$ is the total area of the sky.  Lensing probabilities are shown in Fig.~\ref{lens_prob} for $\feq = 1$.  As in Fig.~\ref{crosssections}, the probability drops with larger cores and larger $\Smin$, and there is a minimum required halo mass $M_i$ for there to be \emph{any} probability of a star being lensed.  The probability rises sharply as $M_i$ increases beyond this minimum, again reflecting the behavior of the individual cross-sections.  However, because of the $n_\mathrm{ucmh}$ factor, the probability falls for larger halos as their increasing rarity begins to dominate over their larger individual cross-sections.  The result is that given some transverse halo velocity $v_t$, there is an initial mass $M_i$ around which a given detection scenario is most sensitive.

\section{Constraints on the primordial curvature perturbation}
\label{sec:powerspectrum}

\begin{figure*}[tb]
\resizebox{\textwidth}!{
\includegraphics{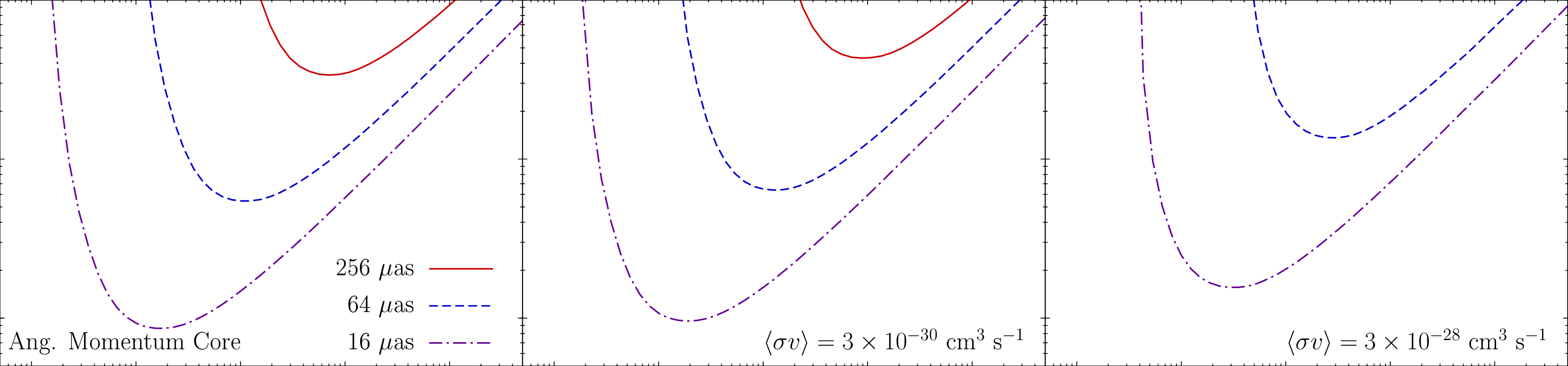}}
\caption{The upper bound on the fraction of dark matter in UCMHs at matter-radiation equality resulting from the nondetection of UCMH lensing events in a survey of 5 million stars.  The solid lines show the upper bound on $\feq$ given a minimum detectable signal of 256 $\mu$as, which corresponds to a 6$\sigma$ detection by Gaia.  The dashed and dotted lines show the upper bounds for more sensitive surveys, with $\Smin=64 \uas$ and $\Smin=16 \uas$, respectively.  As in Fig.~\ref{crosssections}, the three panels show the constraints for three different UCMH density profiles; in the right two panels,  the dark matter particle mass is assumed to be 100 GeV, and the core radius is determined by the ratio $\langle \sigma v\rangle/m_\chi$.}
\label{betaconstraints}
\end{figure*}

The expectation value for the number of observed lensing events is $N_\mathrm{obs} \times P_\mathrm{lensed}$, where $N_\mathrm{obs}$ is the number of stars sampled.  
If a single UCMH lensing event is detected in such a sample, we can invert the Poisson cumulative distribution function to obtain a $95 \%$ confidence lower bound on the lensing probability: $P_\mathrm{lower} = 0.355/N_\mathrm{obs}$.  Conversely, the absence of UCMH lensing events implies a $95 \%$ confidence upper bound on the lensing probability: $P_\mathrm{upper} = 2.996/N_\mathrm{obs}$.  
To determine a feasible value for $N_\mathrm{obs}$, we start with the Gaia astrometric performance as estimated at the time of the Gaia Mission Critical Design Review.\footnote{We calculate the single-epoch precision as 4.3$\times$ the end-of-mission sky-averaged position accuracy in the recommended Gaia model for bright stars -- http://www.rssd.esa.int/index.php?project=GAIA\&page= Science\_Performance
(April 2011).}
  
The Gaia astrometric performance is approximately constant for stars brighter than G=12 (where G is the stellar magnitude in the Gaia passband), and so there is no penalty for including all stars down to that brightness. This cut-off leads to a target list of at least several million stars, based on the G-band all-sky star counts detailed in Gaia Technical Note Gaia\textunderscore ML\textunderscore022 and Ref. \cite{2005ESASP.576..163D}. To ensure a low number of false positives with that number of targets, we require a 6$\sigma$ detection. In this scenario, our canonical value of $S_\mathrm{min}=256 \, \uas$ corresponds to a single-measurement precision of 29 $\mu $as.  This performance is reached by Gaia for a stellar brightness of approximately G=12.5, for which the G-band all-sky star counts predict approximately $7 \times 10^6$ target stars.   We expect some targets to be rejected during the calibration phase because of companions and other effects, so we reduce the estimated $N_\mathrm{obs}$ to $5\times10^6$. With this sample size, the 6$\sigma$ limit for $S_\mathrm{min}$ implies 98\% confidence in a single detection.

In order to translate constraints on the lensing probability into constraints on the initial UCMH mass fraction, we note that $P_\mathrm{lensed}$ has a twofold dependence on $f_\mathrm{eq}$ if $\feq \gsim 1/300$.  The first is a simple linear dependence originating from the number density $n_\mathrm{ucmh} \propto f_\mathrm{eq}$.  The second is a more complex term originating from the growth of the truncation radius $r_t$ as the initial UCMH mass fraction decreases, which increases the post-equality growth factor $g = 1/f_\mathrm{eq}$.  This leads to a weaker restriction on the lensing cross section $A_\mathrm{lensed}$, which must fall completely inside the truncation radius.  However, for the ranges of $S_\mathrm{min}$ and $\deltam$ under consideration, the lensing signal falls off so sharply with increasing star-lens separation that this latter factor is negligible.  We can thus safely assume that $f_\mathrm{eq} \simeq P_\mathrm{lensed}(f_\mathrm{eq}) / P_\mathrm{lensed}(f_\mathrm{eq}=1)$.  If $N_\mathrm{obs}$ stars are monitored and no UCMHs are observed, then we can place an upper bound on $f_\mathrm{eq}$: $f_\mathrm{eq} < (2.996/N_\mathrm{obs})/P_\mathrm{lensed}(f_\mathrm{eq}=1)$.  These upper bounds on $f_\mathrm{eq}$ are shown in Fig.~\ref{betaconstraints}.  We see that a survey of 5 million stars with $S_\mathrm{min} = 256 \mu\mathrm{as}$ can constrain $f_\mathrm{eq}$ to be less than 0.5 for UCMHs with masses between $5 M_\odot \lsim \deltam \lsim 21 M_\odot$ if  $\langle \sigma v\rangle \lsim 3\times10^{-30} \mathrm{cm}^{3}\mathrm{s}^{-1}$.    Figure~\ref{betaconstraints} also shows how future surveys with higher astrometric precisions could improve this constraint.  Even with $S_\mathrm{min} = 16\, \mu\mathrm{as}$, however, we can only constrain $\feq \lsim 0.009$. 

The $f_\mathrm{eq}$ constraints shown in Fig.~\ref{betaconstraints} assume that the inner density profiles of UCMHs are not significantly disturbed from the time of their collapse to the present day.  Since UCMHs have high central densities with steep density profiles and small constant-density cores, it is highly likely that they survive accretion by larger halos \cite{BDE06, BDE08, BDEK10, BSA11}.  However, if $f_\mathrm{eq} \gsim 0.01$, then UCMHs do not grow in isolation before falling into larger, more diffuse halos.  Instead, UCMHs would interact with other UCMHs, and the outcome of such interactions has not been investigated.  A compete analysis of the survival probability for UCMHs with $f_\mathrm{eq} \gsim 0.01$ lies beyond the scope of this work, but we note two reasons to expect that such an analysis will not affect our primary conclusions.  First, UCMHs with $M_i \sim 5\msun$ only generate astrometric microlensing signals greater than 256 $\mu$as if $\beta_y d_l \lsim 5 \times 10^{-4}$ pc.  Given the UCHM's steep density profile, most of the mass beyond this radius could be stripped from the UCMH without significantly changing its astrometric microlensing signature.  Second, we can calculate how UCMH destruction affects our bounds on the primordial power spectrum.  If $s$ is the fraction of UCMHs that survive to the present day, then the bounds on $f_\mathrm{eq}$ shown in Fig.~\ref{betaconstraints} are bounds on the product $s \, f_\mathrm{eq}$.  We will show below that a Gaia-like survey is capable of significantly constraining the amplitude of the primordial power spectrum as long as $s\gsim 0.4$.

We assume that the initial density perturbations are Gaussian and use the Press-Schechter formalism \cite{PS74} to translate the constraints on the initial UCMH mass fraction $f_\mathrm{eq}$ into constraints on the mean-squared amplitude of dark-matter density fluctuations within a sphere of radius $R$ at horizon crossing: $\sigma_{\mathrm{hor}}^2 (R)$.  If we assume that the UCMHs do not accrete prior to matter-radiation equality, then $f_\mathrm{eq}$ equals the fraction of the dark matter contained in regions with dark-matter overdensities greater than the minimum matter overdensity required to form an UCMH ($\deltamin$) and smaller than the matter overdensity required to form a primordial black hole ($\sim\!\!1/4$, corresponding to a radiation overdensity of $1/3$) \cite{Carr75}:
\begin{equation}
f_\mathrm{eq}(\deltam) = \frac{2}{\sqrt{2 \pi \sigma_\mathrm{hor}^2(R)}} \int_{\deltamin}^{1/4} \mathrm{Exp} \left[ -\frac{\delta^2}{2\sigma_\mathrm{hor}^2 (R)}\right] d \delta,
\label{f0integral}
\end{equation}
where $R$ is the comoving radius containing a dark matter mass $\deltam$ [see Eq.~(\ref{delm})].   We are primarily interested in values of $\sigmahor^2$ that are much less than 1/4, so $f_\mathrm{eq}$ is insensitive to the upper limit on this integration.  In contrast, $f_\mathrm{eq}$ depends very strongly on the value of $\deltamin$.

\begin{figure*}[tb]
\resizebox{\textwidth}!{
\includegraphics{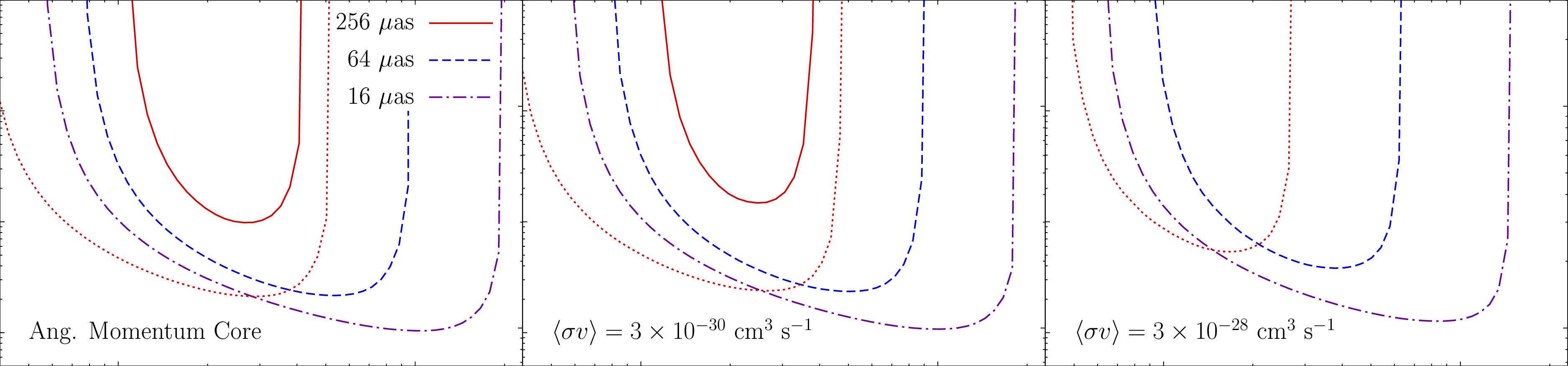}}
\caption{The bounds on the primordial perturbation amplitude resulting from a survey of 5 million stars.  The left axis gives the amplitude of the primordial curvature power spectrum, while the right axis gives the root-mean-squared amplitude of dark matter density fluctuations within a comoving sphere of radius $R=k^{-1}$ at horizon entry.  The solid curves show the upper bounds that would follow from the failure to detect lensing events at different values of $\Smin$.  The dotted curves shows the lower bound that would result if an UCMH lensing event were detected by a survey with $\Smin = 256 \uas$.  The three panels are the same as the three panels in Fig.~\ref{betaconstraints}.}
\label{pkconstraints}
\end{figure*}

Several studies of UCMHs assume that $\deltamin \simeq 10^{-3}$ for all UCMH masses \cite{RG09, SS09, JG10}, but this value for $\deltamin$ is only correct to an order of magnitude, and it ignores the scale-dependence of $\deltamin$.  Since smaller-scale perturbations enter the Hubble horizon prior to larger-scale perturbations, they have more time to grow and reach the collapse threshold before a given collapse redshift.  Therefore, if we define $\deltamin(R)$ to be the overdensity at horizon crossing required for the region to collapse prior to a given redshift, $\deltamin$ will decrease slightly as $R$ decreases.  A precise calculation of $\deltamin$ as a function of scale and collapse redshift was recently presented in Ref.~\cite{BSA11}, and we briefly summarize their result in the Appendix.  Given $\deltamin(R)$ and the upper bounds on $\feq(\deltam)$ shown in Fig.~\ref{betaconstraints}, we can use Eq.~(\ref{f0integral}) to derive upper bounds on $\sigma_{\mathrm{hor}}^2 (R)$.  These upper bounds are shown in Fig.~\ref{pkconstraints} (see right axis).  We also show the lower bound on $\sigma_{\mathrm{hor}}^2$ that would result from the detection of one UCMH lensing event by Gaia with $S_\mathrm{min}=256\uas$.

The translation of constraints on $\sigma_{\mathrm{hor}}^2(R)$ to constraints on the primordial power spectrum ${\cal P}_{\cal R}(k)$ is model-dependent; it depends on the scale-dependence of ${\cal P}_{\cal R}(k)$.  We assume that ${\cal P}_{\cal R}$ is locally scale-invariant (i.e. that it does not vary significantly when $k$ changes by a factor of a few), and we constrain its amplitude at different scales.  With this assumption, \mbox{$\sigma_{\mathrm{hor}}^2(R)  \simeq 0.908 {\cal P}_{\cal R}\left(k=R^{-1}\right)$} \cite{BSA11}; a brief derivation of this relation is given in the Appendix.  

The left axis of Fig.~\ref{pkconstraints} shows the upper bounds on ${\cal P}_{\cal R}(k)$ that follow from the absence of UCMH lensing signals in a survey of 5 million stars.  A Gaia-like survey with $\Smin = 256 \uas$ probes the primordial power spectrum in a fairly narrow band around $k \sim 2500$ Mpc$^{-1}$.  If the core radius of the UCMH is sufficiently small, Gaia could set an upper bound ${\cal P}_{\cal R} \lsim 10^{-5}$ at these scales.  Fig.~\ref{pkconstraints} also shows how surveys with higher astrometric precision could further strengthen this constraint over a wider range of scales.  Alternatively, if Gaia observes a lensing signal from an UCMH, we can put a lower bound on ${\cal P}_{\cal R}$; this lower bound also is shown in Fig.~\ref{pkconstraints}.

As illustrated in Fig.~\ref{pkconstraints}, the ${\cal P}_{\cal R}$ constraints derived from astrometric microlensing depend on the properties of the dark matter particle; the upper bounds on ${\cal P}_{\cal R}$ increase as $\langle \sigma v \rangle/m_\chi$ increases because the lensing signal is smaller for UCMHs with larger core radii.  The only constraints on ${\cal P}_{\cal R}$ for $k\gsim 10$ Mpc$^{-1}$ that do not depend on the properties of the dark matter particle are the bounds from the null detection of PBHs 
(${\cal P}_{\cal R} \lsim 0.05$ at $k\simeq 3000$ Mpc$^{-1}$ \cite{JGM09}) and CMB spectral distortions (assuming local scale invariance, ${\cal P}_{\cal R} \lsim 2\times10^{-5}$ at $k\simeq 3000$ Mpc$^{-1}$ \cite{JEB12}).
For small UCMH cores, the constraint from Gaia ($\Smin = 256 \uas$) is twice as strong than the bound from CMB spectral distortions and over three orders of magnitude stronger that the bounds from PBHs!

We now revisit the issue of UCMH survival.  As long as $s f_\mathrm{eq} \lsim 0.1$, decreasing $s$ has little impact on the upper bounds on ${\cal P}_{\cal R}(k)$ shown in Fig.~\ref{pkconstraints}, which were calculated assuming that all UCMHs survive ($s=1$).  For example, if $s=1$, no UCMH detections in a survey of 5 million with $\Smin = 16 \uas$ implies ${\cal P}_{\cal R} < 1.1\times10^{-6}$ for $k=8800$~Mpc$^{-1}$; if $s=0.1$, this upper bound increases to ${\cal P}_{\cal R} < 2.5\times10^{-6}$.  The constraints are more sensitive to changes in $s$ for larger values of $s f_\mathrm{eq}$; setting $s=0.1$ for a survey with $\Smin = 64 \uas$ increases the upper bound on ${\cal P}_{\cal R}$ at $k=5000$ Mpc$^{-1}$ by an order of magnitude.  For a Gaia-like survey with $\Smin = 256 \uas$, setting $s=0.4$ implies ${\cal P}_{\cal R} < 2.3\times10^{-4}$ for $k=2600$ Mpc$^{-1}$.  While weaker than the upper bound shown in Fig.~\ref{pkconstraints}, this upper bound on ${\cal P}_{\cal R}$ is still significantly stronger than the constraint derived from the absence of PBHs. 
 
\section{UCMHs as gamma-ray sources}
\label{sec:gammarays}

\begin{figure*}
\resizebox{\textwidth}!{
\includegraphics{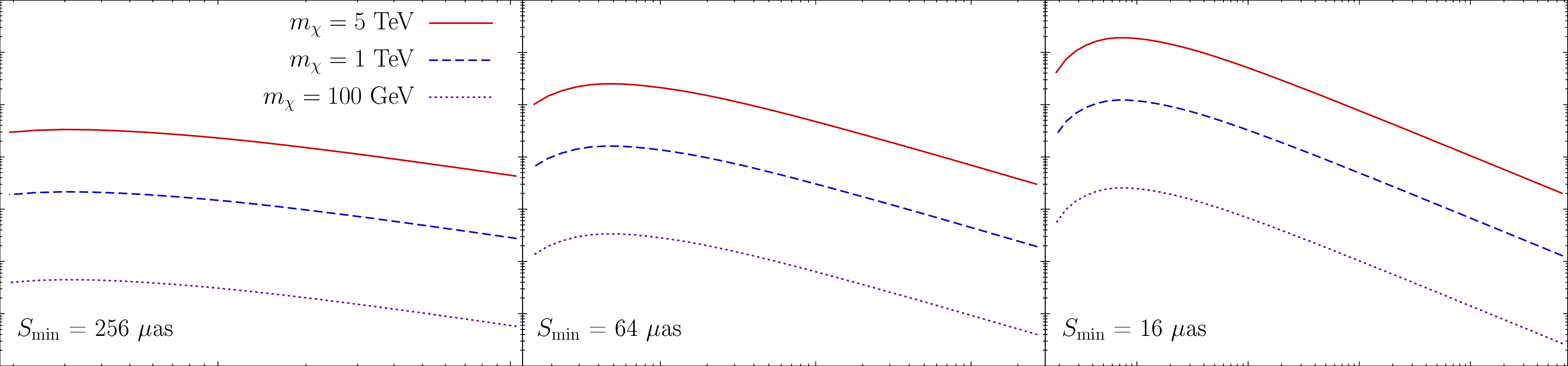}}
\caption{The value of $\sigv$ below which constraints on $\feq$ from astrometric microlensing are more stringent than the constraints from gamma-ray emission.  The three panels show different values of $\Smin$, and in all cases, we assume that the astrometric microlensing constraints are derived from a survey of 5 million stars that did not detect lensing events from UCMHs.   The three curves in each panel show the results for different dark matter particle masses.}
\label{sigmavlimits}
\end{figure*}

While they are a significant improvement over the constraints from PBHs, the upper bounds on ${\cal P}_{\cal R}(k)$ shown in Fig.~\ref{pkconstraints} are all weaker than the bounds derived from the fact that Fermi-LAT satellite has not observed gamma-rays from dark matter annihilation inside UCMHs, but these constraints depend on the properties of the dark matter particle.  Ref.~\cite{BSA11} assumed that $\mdm = 1$ TeV, $\sigv = 3\times10^{-26}$ cm$^3$ s$^{-1}$, and that all the dark matter particles annihilate to $b\bar{b}$ pairs.  With these assumptions, they showed that the most stringent bounds on the UCMH population in the mass range \mbox{$3\times10^{-6} M_\odot \lsim \deltam \lsim 3\times10^{5}M_\odot$} follow from Fermi-LAT's failure to detect individual UCMHs within our galaxy as gamma-ray point sources.  The resulting constraint on the primordial power spectrum is ${\cal P}_{\cal R} \lsim 3\times10^{-7}$ for $k\simeq 3000$ Mpc$^{-1}$.  Constraints on ${\cal P}_{\cal R}$ from astrometric microlensing by UCMHs probe a different region of dark matter parameter-space; as the ratio $\langle \sigma v \rangle/m_\chi$ decreases, the Fermi-LAT constraints get weaker, while the constraints from astrometric microlensing either get stronger or are unaffected.  In this section, we find the region of dark matter parameter-space in which the astrometric microlensing constraints on ${\cal P}_{\cal R}$ surpass the constraints from Fermi-LAT.  We will see that the constraints shown in the left panel of Fig.~\ref{pkconstraints} can be considered conservative; if the dark matter annihilation leads to larger UCMH core radii, then Fermi-LAT places a stronger upper bound on ${\cal P}_{\cal R}$.

An UCMH located at a distance $d$ produces an observed gamma-ray flux
\beq
{\cal F} =  \frac{\sigv}{2d^2 \mdm^2} \left( \int_{E_\mathrm{th}}^{m_\chi} \frac{\drm N}{\drm E}\drm E\right)\left(\int_0^{r_t} r^2 \rho^2(r) \drm r\right),
\label{flux}
\eeq
where $E_\mathrm{th}$ is the minimum energy required for detection, $r_t$ is the radius of the UCMH [see Eq.~(\ref{truncationradius})], and we have assumed that the annihilation proceeds via a single channel with a differential photon yield given by ${\drm N}/{\drm E}$.  For our assumed UCMH density profile [Eq.~(\ref{coredprofile})] with $r_t \gg r_c$,
\beq
\int_0^{r_t} r^2 \rho^2(r) \drm r \simeq \frac{16}{105} \rho_0^2 r_c^3.
\label{volint}
\eeq
Since $r_t \gg r_c$ for all UCMHs, we see that the flux only depends the UCMH's core radius and density.  If the core radius of the UCMH is set by dark matter annihilation, then $\rho_0$ depends only on the properties of the dark matter particle, but $r_c \propto M_i^{1/3}$; it follows that $\rho_0^2 r_c^3 \propto M_i(\mdm/\sigv)^{2/3}$.  Conversely, if the annihilation core is smaller than the core produced by nonradial infall, then $\rho_0^2 r_c^3 \propto M_i^{1.09}$.

Following Ref.~\cite{BSA11}, we define $d_\mathrm{obs}$ to be the distance at which the gamma-ray flux from an UCMH equals $4\times10^{-9}$ photons cm$^{-2}$ s$^{-1}$ with $E_\mathrm{{th}} = 100$ MeV; this is the flux required for a 5$\sigma$ detection of a point-source after one year of observations by Fermi-LAT.  If all of the dark matter is presently contained in UCMHs ($\feq \gsim 1/300$), then the expected number of UCMHs that Fermi-LAT should detect is $\feq M_{d<d_\mathrm{obs}}/ \deltam$, where $M_{d<d_\mathrm{obs}}$ is the mass of the dark matter enclosed in a sphere with radius $d_\mathrm{obs}$ centered at our location.  Since Fermi-LAT has not detected any UCMHs after one year of operation, we can conclude that  \mbox{$\feq < 2.996 {\deltam}/{M_{d<d_\mathrm{obs}}}$} at 95\% confidence.  If the constraint derived from astrometric microlensing is $\feq< f_\mathrm{eq,\mathrm{AM}}$, then the astrometric microlensing constraints are more powerful than the constraints from Fermi-LAT if 
\beq
d_\mathrm{obs} < \left(\frac{3\times 2.996 \deltam}{4\pi f_\mathrm{eq,\mathrm{AM}} \rho_\mathrm{dm,0}}\right)^{1/3}.
\label{dobslimit}
\eeq
All the upper bounds on $\feq$ shown in Fig.~\ref{betaconstraints} require $d_\mathrm{obs} < 38$ pc to surpass the Fermi constraints, so our assumption that the dark matter density is uniform within a sphere with radius $d_\mathrm{obs}$ is justified.  

Since $d_\mathrm{obs}$ depends on the properties of the dark matter particle, the upper bound on $d_\mathrm{obs}$ given by Eq.~(\ref{dobslimit}) implies an upper limit on $\sigv$ for a given value of $\mdm$.  These upper bounds on $\sigv$ are shown in Fig.~\ref{sigmavlimits}.  To derive these limits, we used values for $d_\mathrm{obs}(M_i)$ calculated using an extended version  of DarkSUSY \cite{DarkSUSY,SS09},
assuming all the dark matter particles annihilate to $b\bar{b}$ pairs, for three values of $\sigv$ and $\mdm$ \cite{PatEmail}.\footnote{These values for $d_\mathrm{obs}$ were calculated using a different UCMH profile: $\rho (r) = \rho_0$ for $r < r_c$ and Eq.~(\ref{uncoredprofile}) for $r > r_c$.  With this density profile, the integral given by Eq.~(\ref{volint}) equals $\rho_0^2 r_c^3$.  To obtain $d_\mathrm{obs}$ for our UCMH density profile [Eq.~(\ref{coredprofile})], we reduce these $d_\mathrm{obs}$ values by a factor of $\sqrt{16/105}$ before using them in our analysis.}
Since $d_\mathrm{obs} \propto {\cal F}^{1/2}$, $d_\mathrm{obs} \propto \sigv^{1/6}$ if the UCMH core is set by dark matter annihilation, and $d_\mathrm{obs} \propto \sigv^{1/2}$ if the UCMH core is set by the nonradial infall; we used these scalings to obtain $d_\mathrm{obs}(M_i, \sigv)$ functions for each value of $\mdm$.   We then found the value of $\sigv$ that saturates Eq.~(\ref{dobslimit}) given the upper bounds $f_\mathrm{eq,\mathrm{AM}} (M_i)$ shown in Fig.~\ref{betaconstraints}.

Figure~\ref{sigmavlimits} shows that the dark matter particle must be weakly self-annihilating, with $\sigv \ll 3\times10^{-26}$ cm$^3$ s$^{-1}$, if searches for astrometric microlensing by UCMHs are to yield stronger constraints on $\feq$ than Fermi-LAT.  For these small values of $\sigv/\mdm$, the core radius set by dark matter annihilation is smaller than the core radius set by the breakdown of radial infall.  Therefore, as long as $\sigv$ and $\mdm$ remain unknown, we can safely assume that the UCMH density profile is not affected by dark matter annihilation when calculating the bounds on $\feq$ and ${\cal P}_{\cal R}$ that would result from the failure of an astrometric survey to observe UCMH lensing events.  Such bounds are, in effect, the worst-case scenario; although they would be weakened if dark matter within UCMH annihilates sufficiently to increase the core radius, that rate of dark matter annihilation would be sufficient to make UCMHs detectable by Fermi-LAT.

\section{Summary and Discussion}
\label{sec:summary}
UCMHs are small dark matter minihalos that form shortly after matter-radiation equality ($z\gsim 1000$).  Since they form earlier than standard minihalos, UCMHs have steep density profiles ($\rho \propto r^{-9/4}$) and high central densities.  These properties make UCMHs ideal targets for detection by astrometric microlensing \cite{EL11}: when an UCMH passes in front of a star, the light from that star is gravitationally deflected, and the star's image follows a distinctive trajectory.  We calculated the astrometric microlensing signatures generated by UCMHs, and we found that the image trajectories depend strongly on the density profile near the center of the UCMHs.  We assume that UCMHs have a constant-density core; the minimal core radius is determined by the breakdown of the radial infall approximation that yielded the $\rho \propto r^{-9/4}$ density profile, but the core radius may be increased if dark-matter annihilations limit the UCMH central density.   Increasing the core radius of an UCMH decreases its astrometric microlensing signature, which implies that astrometric microlensing searches for UCMHs are most effective if the dark-matter annihilation rate is suppressed. 

The image trajectories produced by microlensing with UCMHs are arc-shaped; the star's image moves slowly as the center of the UCMH approaches, then it rapidly traverses the arc as the center of the UCMH passes by, and finally it slows down at the arc's opposite end as the UCMH center moves away.  Since UCMHs are extended objects, it takes several decades for the star's image to return to its true position, which makes UCMH microlensing events easily distinguishable from lensing by point masses.  The minimal motion of the image prior to the UCMH center's passage prompts us to adopt the same observing strategy as Ref. \cite{EL11}: the first two years of observations are used to measure the stars' proper motions and parallaxes, and then astrometric microlensing events over the next four years are detected as deviations from the image trajectory predicted from these measurements.  We find that a Gaia-like survey, with a per-epoch astrometric precision of 29 microarcseconds for 5 million target stars, is most sensitive to UCMHs with initial masses of $\sim\!\!7\, \msun$.   Surveys with higher astrometric precisions can constrain the abundance of slightly smaller UCMHs, but one-microarcsecond precision is required to detect astrometric microlensing by UCMHs with masses less than $\sim\!\!0.01\, \msun$.  No current or planned astrometric instrument can deliver such precision for large numbers of targets.  Therefore, at least for the near future, astrometric microlensing cannot be used to search for UCMHs that result from the enhancement of density perturbations during QCD phase transition \cite{1999PhRvD..59d3517S, SS09} or during a matter-dominated era prior to big bang nucleosynthesis \cite{ES11}.  Astrometric microlensing can detect UCMHs that form from large density fluctuations with wavelengths between 0.6~kpc and 6~kpc.

Since UCMHs form from large-amplitude density fluctuations ($\delta \rho/\rho \gsim 10^{-3}$ at horizon entry), their abundance probes the amplitude of the primordial power spectrum on these small scales.  If dark matter self-annihilates, the high density of dark matter within UCMHs makes them bright gamma-ray sources, and they may be detected by Fermi-LAT \cite{SS09, JG10, BSA11}.  The fact that Fermi-LAT has not detected gamma-ray emission from UCMHs can be used to constrain the primordial power spectrum, but this constraint depends on the annihilation rate of dark matter particles within UCMHs \cite{BSA11}.  The limits on the abundance of UCMHs derived from searches for their astrometric microlensing signatures would complement the constraints derived from the nondetection of gamma-ray emission from UCMHs because the UCMHs produce larger image deflections if the dark-matter annihilation rate is low.   If dark matter annihilation within UCMHs is efficient enough to change the UCMH density profile and weaken the constraints from astrometric microlensing, then the strongest bounds on the power spectrum follow from the fact that Fermi-LAT has not yet observed emissions from UCMHs .  However, if dark matter annihilation does not increase the size of the UCMHs' cores, we find that a search for UCMHs by Gaia could constrain the amplitude of the primordial power spectrum to be less than 10$^{-5}$ on scales of \mbox{$k\simeq2700$ Mpc$^{-1}$}, an improvement of three orders of magnitude over the bound derived from the absence of PBHs \cite{JGM09}.  Unlike other constraints derived from UCMHs, this upper bound on the primordial power spectrum does not depend on the properties of the dark matter particle. 

When calculating the bounds on the primordial power spectrum that result from limits on the local UCMH abundance, we assumed that the UCMHs' innermost regions survive to the present day.  UCMHs are probably not disrupted during their accretion by larger halos because they have small cores with high central densities \cite{BDE06, BDE08, BDEK10, BSA11}.  The outcome of interactions between UCMHs is less certain, and the bound on the local UCMH abundance derived from astrometric microlensing searches is not low enough to ensure that such interactions are uncommon.  If UCMH-UCMH interactions increase the UCMHs' core radii or decrease their central densities, then our projected bounds on the primordial power spectrum are weakened.  We encourage further study of UCMH survival in scenarios in which UCMH interactions are common to determine exactly how UCMHs are effected by such interactions and what impact this has on their astrometric microlensing signatures. We note, however, that Gaia could significantly improve the PBH constraint on the primordial power spectrum even if up to 60\% of UCMHs that formed initially cannot generate an observable astrometric microlensing signal today.   

Our analysis could also be refined by considering more sophisticated techniques for detecting astrometric lensing events.  Our detection strategy is designed to evaluate the level of astrometric signal produced by UCMHs.  An optimized detection strategy, perhaps based on matched filters, would provide greater sensitivity and improved false positive rejection.  Finally, we note that gravitational lensing by dark matter minihalos will also generate time delays, which could be detected using high-precision pulsar timing \cite{SHF07, BAZ11}; it would be interesting to investigate how pulsar timing may constrain the abundance of UCMHs.

The upper limit on the amplitude of the primordial power spectrum on small scales derived from the absence of PBHs is a valuable tool in our quest to understand inflationary physics \cite{1993PhRvD..48..543C,2000PhRvD..62d3516L,2008JCAP...04..038K,2008JCAP...07..024P,2010PhRvD..82d7303J, 2011arXiv1107.1681L, 2011arXiv1112.5601B}.   Limits on the UCMH abundance can provide a much stronger bound on the amplitude of the primordial power spectrum on small scales.  Currently, the only constraints on the UCMH population are derived from their possible emission of dark matter annihilation products; these constraints do not apply if the dark matter annihilation rate is sufficiently suppressed.  We have shown that astrometric microlensing offers a promising alternate method of detecting UCMHs and thus constraining inflationary physics.

\acknowledgments
We thank Pat Scott for providing the $d_\mathrm{obs}$ values used in Section \ref{sec:gammarays} and for several useful discussions and comments on our manuscript.  We also thank Torsten Bringmann for his guidance on the calculation of $\delta_\mathrm{min}$.  FL acknowledges support from the Canadian Institute for Theoretical Astrophysics.  ALE is supported by NSERC, the Perimeter Institute for Theoretical Physics and the Canadian Institute for Advanced Research.  Research at the Perimeter Institute is supported by the Government of Canada through Industry Canada and by the Province of Ontario through the Ministry of Research and Innovation.

\appendix
\section{From $f_\mathrm{eq}(M_i)$ to $P_{\cal R}(k)$}
\label{sec:appendix}
In this Appendix, we briefly review how bounds on $f_\mathrm{eq}$ translate into bounds on the amplitude of the primordial power spectrum $P_{\cal R}(k)$.  From Eq.~(\ref{f0integral}), we see that we must first calculate the minimum matter overdensity required to form an UCMH ($\deltamin$).  Once we have $\deltamin(R)$, we can find the value of $\sigma_{\mathrm{hor}}^2 (R)$ that yields the desired value of $f_\mathrm{eq}$.  If we then assume a specific scale dependence for $P_{\cal R}(k)$, we can use
$\sigma_{\mathrm{hor}}^2 (R)$ to obtain $P_{\cal R}(k)$.  This procedure is described in detail in Ref. \cite{BSA11} for several models of $P_{\cal R}(k)$.  We briefly summarize the key results here.

During matter domination, a region collapses when the linear overdensity reaches the collapse overdensity: $\delta_\mathrm{coll} = (3/5)(3\pi/2)^{2/3}\simeq 1.686$.  The amplitude of subhorizon density perturbations during the matter-dominated era is proportional to the amplitude of the curvature perturbation ${\cal R}$ at horizon entry:
\beq
\delta_\chi(k,z) = \frac{2}{5}  \frac{k^2}{H_0^2\Om } T(k){\cal D}(z) {\cal R}(k)
\label{mdpert}
\eeq
where $\Om$ is the present-day matter density divided by the critical density, $T(k)$ is the transfer function normalized to unity on large scales, and 
\beq
{\cal D}(z) = \frac{(1+z)^2}{(1+\zeq)^3} 
\left[2+\left(\frac{1+\zeq}{1+z}-2\right)\sqrt{1+\frac{1+\zeq}{1+z}}\right]^2
\eeq
is the growth function after matter-radiation equality (e.g. \cite{WeinbergCosmo}).  We can use Eq.~(\ref{mdpert}) to find the curvature perturbation required to make \mbox{$\delta_\chi(k,z_\mathrm{coll})=\delta_\mathrm{coll}$} at some collapse redshift $z_\mathrm{coll}$.  We follow Ref.~\cite{BSA11} and set $z_\mathrm{coll}=1000$.   

Next, we need to calculate the dark matter density perturbation at horizon entry that corresponds to this curvature perturbation; this is $\deltamin$.  The overdensity at horizon entry is a gauge-dependent quantity, however.  Since $\delta_\mathrm{coll}$ was evaluated in the density rest frame (total matter gauge), and the relevant modes for UCMHs entered the horizon during radiation domination, Ref.~\cite{BSA11} evaluates $\deltamin$ in the rest frame of the radiation.  In this gauge, the dark matter density perturbation during radiation domination is given by
\beq
\delta_\chi(k,a) = 6{\cal R}(k)\left[\ln \theta +\gamma_E-\frac{1}{2}-\mathrm{Ci}(\theta)+\frac{\sin \theta}{2\theta}\right]
\label{rdpert}
\eeq
where $\gamma_E$ is the Euler-Mascheroni constant, Ci is the cosine integral function, and $\theta=k/(\sqrt{3}aH)$.   Evaluating this expression at horizon entry \mbox($k=aH$) and setting ${\cal R}$ to the minimum value required to make \mbox{$\delta_\chi(k,z_\mathrm{coll})=\delta_\mathrm{coll}$} yields
\beq
\deltamin(k,z_\mathrm{coll}) = \frac{5}{6}\delta_\mathrm{coll} \frac{H_0^2 \Om}{k^2 T(k)}\frac{0.988}{D(z_\mathrm{coll})}.
\eeq
For $k=5.1\times 10^{4}$ Mpc$^{-1}$, which corresponds to $\deltam = 10^{-3}\msun$, $\deltamin = 2.2\times10^{-3}$.   As expected, $\deltamin$ is slightly larger for larger UCMHs; for $\deltam = 1000\msun$, $\deltamin = 3.4\times10^{-3}$.

The primordial power spectrum determines $\sigma^2 (R,t)$:
\beqa
\sigma^2 (R,t) &=& \int_0^\infty F^2(kR){\cal P}_\chi(k,t) \frac{\drm k}{k}, \nonumber\\
&=&\frac{1}{9}\int_0^\infty F^2(kR) \frac{k^4}{a^4H^4}T_\chi^2(\theta) {\cal P}_{\cal R}(k)  \frac{\drm k}{k},
\eeqa
where $F(x)= 3x^{-3}(\sin x - x \cos x)$ is the Fourier transform of the top-hat window function, and \mbox{$T_\chi(\theta) \equiv \delta_\chi/(\theta^2 \cal{R})$} with $\delta_\chi$ given by Eq.~(\ref{rdpert}).  If we evaluate $\sigma^2 (R,t)$ at horizon crossing [$R=(aH)^{-1}$] and change the integration variable to $x\equiv kR$, we have
\beq
\sigma_{\mathrm{hor}}^2(R) =\frac{1}{9}\int_0^\infty x^3 T_\chi^2\left(\theta=\frac{x}{\sqrt{3}}\right)  F^2(x) {\cal P}_{\cal R}\left(k=\frac{x}{R}\right) \drm x
\eeq
The factor $x^3 T_\chi^2 F^2$ serves as a modified window function: it is peaked at $kR \simeq 2.4$, and if ${\cal P}_{\cal R}(k)$ is scale-invariant, 80\% of the integral's value comes from integrating the range $1<kR<4$.  We assume that ${\cal P}_{\cal R}$ is nearly constant within this limited range of scales.  In this case, ${\cal P}_{\cal R}(k)$ can be pulled outside the integral, and we are left with \mbox{$\sigma_{\mathrm{hor}}^2(R)  \simeq 0.908 {\cal P}_{\cal R}\left(k=R^{-1}\right)$}.

\end{document}